# Quasiparticle and Excitonic Structures of Few-layer and Bulk GaSe: Interlayer Coupling, Self-energy, and Electron-hole Interaction

Fanhao Jia[1,2,6], Zhao Tang[3,4,6], Greis J. Cruz[3], Weiwei Gao[5], Shaowen Xu[2], Wei Ren[2†], and Peihong Zhang[3]*

[1]*Department of Physics, Hangzhou Dianzi University, Hangzhou 310018, China*
[2]*International Centre of Quantum and Molecular Structures, Department of Physics, Shanghai University, Shanghai 200444, China*
[3]*Department of Physics, University at Buffalo, State University of New York, Buffalo, New York 14260, USA*
[4]*Center for Computational Materials, Oden Institute for Computational Engineering and Sciences, The University of Texas at Austin, Austin, Texas 78712, USA*
[5]*Key Laboratory of Material Modification by Laser, Ion, and Electron Beams, Dalian University of Technology, Ministry of Education, Dalian 116024, China*
[6]*FJ and ZT contributed equally to this work.*
[†]Emails: renwei@shu.edu.cn
*Emails: pzhang3@buffalo.edu

## Abstract

Metal monochalcogenide GaSe is a classic layered semiconductor that has received increasing research interest due to its highly tunable electronic and optical properties for ultrathin electronics applications. Despite intense research efforts, a systematic understanding of the layer-dependent electronic and optical properties of GaSe remains to be established, and there appear significant discrepancies between different experiments. We have performed GW plus Bethe-Salpeter equation (BSE) calculations for few-layer and bulk GaSe, aiming at understanding the effects of interlayer coupling and dielectric screening on excited state properties of GaSe, and how the electronic and optical properties evolve from strongly two-dimensional (2D) like to intermediate thick layers, and to three-dimensional (3D) bulk character. Using a new definition of the exciton binding energy, we are able to calculate the binding energies of all excitonic states. Our results reveal an interesting correlation between the binding energy of an exciton and the spread of its wave function in the real and momentum spaces. We find that the existence of (nearly) parallel valence and conduction bands facilitates the formation of excitonic states that spread out in the momentum space. Thus, these excitons tend to be more localized in real space and have large exciton binding energies. The interlayer coupling substantially suppresses the Mexican-hat-like dispersion of the top valence band seen in monolayer system, explaining the greatly enhanced photoluminescence (PL) as layer thickness increases. Our results also help resolve apparent discrepancies between different experiments. After including the quasiparticle and excitonic effects as well the optical activities of excitons, our results compare well with available experimental results.

## I. Introduction

Gallium selenide (GaSe) belongs to a large family of post-transition metal mono-chalcogenide semiconductors [1-3] which has emerged as an important class of layered materials for applications ranging from ultrathin optoelectronics [4,5] to photocatalysis . Bulk GaSe has an optical band gap of around 2 eV at room temperature (~2.1 eV at 10 K) [6-8]. Unlike monolayer transition metal dichalcogenide such as $MoS_2$, which has a direct band gap, monolayer GaSe has an indirect band gap. Interestingly, the nature of the band gap (i.e., direct or indirect) of bulk GaSe is still not settled (we will discuss this issue later). The band dispersion near the valence band maximum (VBM) of few-layer GaSe also has an interesting Mexican-hat-like (MH-like) shape [5,9,10], with the VBM slightly shifted from the Γ-point, while the conduction band minimum (CBM) is always at the Γ-point.

It is well-known that the band gaps (both quasiparticle and optical) of layered materials often increase significantly as the number of layers decreases due to quantum confinement, reduced interlayer coupling, and dielectric screening effects. For example, it was reported that the energy of photoluminescence (PL) A peak of InSe increases from 1.27 eV for the bulk phase to 1.88 eV (bilayer) and 2.60 eV (monolayer) [11], suggesting a widely tunable optical gap by controlling the layer thickness. GaSe is expected to have similar tunability. However, experimental reports have yielded seemingly inconsistent findings. Earlier experiments suggest that the optical band gap of GaSe does not depend sensitively on the layer thickness [12-14]. More recent experiments, on the other hand, do observe significant layer dependence of the PL spectra. The peak position increases from 1.99 eV (bulk) to 2.42 eV (bilayer), and no PL peaks were observed for the monolayer system [15]. Cathodoluminescence (CL) experiments, in contrast, show a broad peak at about 3.3 eV [16] for the monolayer GaSe. Quasiparticle (QP) band gaps measured by scanning tunneling spectroscopy (STS) are 3.5 eV, 3.0 eV, and 2.3 eV for monolayer, bilayer, and trilayer GaSe, respectively [17]. Therefore, there appear to be significant discrepancies between experiments.

In this work, we carried out accurate many-body perturbation theory (MBPT) calculations within the GW plus Bethe-Salpeter equation (BSE) approach [18,19] to understand the quasiparticle and optical properties of bulk and few-layer GaSe. The GW method is one of the most accurate methods for calculating QP excitations, whereas the excitonic and optical properties can be predicted by solving the BSE. We find that, even in this moderate-gap material, some above-the-gap bulk exciton states can be highly localized and have surprisingly large binding energies (as large as 168 meV) thanks to several pairs of nearly parallel valence and conduction bands in portion of the Brillouin zone (BZ), which lead to delocalization of the exciton wave functions in the *k*-space and localization in real space. The calculated exciton binding energy of the lowest excitonic state of bulk GaSe, after extrapolating to the converged value, agrees very well with experimental result. The unique excitonic structures and exciton binding energies of these systems are analyzed in detail to illustrate the effects of interlayer coupling and layer-dependent dielectric screening on excited states properties of GaSe. Our work also helps reconcile the seemingly inconsistent experimental results.

## II. Computational details

The structures of few-layer and bulk GaSe were optimized using the meta-generalized gradient approximation (meta-GGA) within the strongly constrained and appropriately normed (SCAN) functional plus revised Vydrov-van Voorhis nonlocal correlation $r$VV10 [20], which is implemented in Vienna *ab initio* simulation package (VASP) [21]. The SCAN+rVV10 functional has been shown to accurately predict the structural properties of a range of layered materials [20]. A 25 Å vacuum slab was added in few-layer systems and a slab-truncated Coulomb potential [22] was used to minimize the interaction between periodic adjacent layers.

We used a local version of the BerkeleyGW code [23] to perform the GW [18] and BSE [19] calculations. The recently developed acceleration methods [24-27] were used for the GW calculations. These methods lead to a combined speed-up factor of up to 1,000 for GW calculations for two dimensional (2D) materials compared with conventional band-by-band summation and uniform $k$-point sampling approaches. The Hybertsen-Louie generalized plasmon-pole (HL-GPP) model [18] was used to extend the static dielectric function to finite frequencies. The density functional theory (DFT) part of the MBPT calculation was carried out using the Perdew-Burke-Ernzerhof (PBE) [28] functional and the Troullier–Martins norm-conserving pseudopotentials [29]. A plane wave cutoff energy of 60 Ry was used in the DFT calculations, and we used a high cutoff energy of 30 Ry for the dielectric matrices and the *e-h* kernel in the GW and BSE calculations.

A dual-grid method [19] was applied to reduce the workload of the BSE calculations. For the few-layer systems, the *e-h* kernel was first calculated on an 18×18×1 coarse $k$-grid; the results were then interpolated onto a finer grid (96×96×1 for the monolayer, 90×90×1 for the bilayer, and 78×78×1 for the trilayer) while solving the BSE for the excitonic and optical properties. For the bulk system (which has a rather thick unit cell, as we will discuss later), the *e-h* kernel was calculated on a 12×12×2 $k$-grid, which was then interpolated onto a 36×36×8 fine grid in solving the BSE. We have carefully checked the convergence of our GW and BSE results; see Supplemental Material at [30] Fig. S1 for the details of the convergence tests.

## III. Results and discussion

### A. Structural properties of few-layer and bulk GaSe

Bulk GaSe exhibits several polymorphic modifications with different stacking and numbers of layers, while the $\varepsilon$-GaSe polytype (space group of $P\bar{6}m2$) is the most common phase at room temperature [31]. The structure ε-GaSe consists of two hexagonal layers (Fig. 1) with Ga assuming a tetrahedral coordination and Se forming three covalent bonds with Ga. The hexagonal layer, although consisting of multiple atomic planes, is tightly bonded; therefore, they are considered a single layer (monolayer) in our study (with a layer thickness $d_1$ as shown in Fig.1). We assume the ε-like stacking in our study of few-layer and bulk GaSe, as shown in Fig. 1.

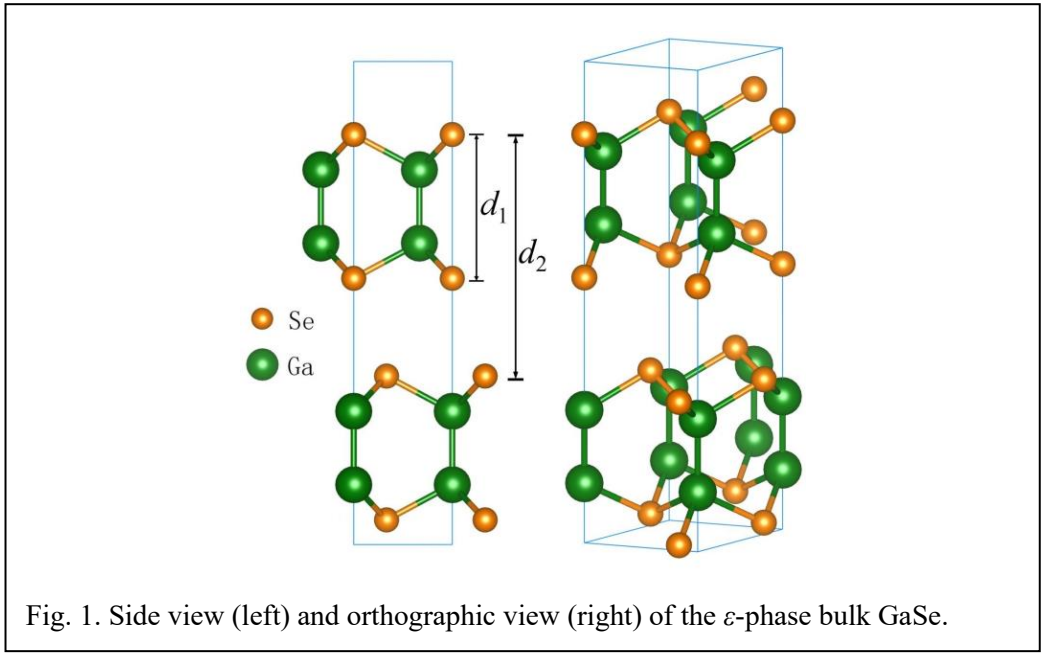

Fig. 1. Side view (left) and orthographic view (right) of the ε-phase bulk GaSe.

Table I compares the optimized structure of few-layer GaSe and the experimental structure of bulk GaSe [32]. The optimized lattice constant $a$ and layer thickness $d_1$ (defined in Fig. 1) of the few-layer systems show little change with the increasing number of layers and are very close to the experimental values (within 0.5%), suggesting the accuracy of the theory and the weak interlayer coupling effects on the structure properties. The $d_2$ parameter is $d_1$ plus the interlayer spacing (see Fig. 1).

Table I. Lattice parameters (in Å, defined in Fig.1) of few-layer and bulk GaSe used in this study. The values of few-layer systems are calculated using the SCAN+$r$VV10 functional, while those of the bulk phase are taken from experiments [32].

|  | 1L | 2L | 3L | Bulk (exp) |
|---|---|---|---|---|
| $a$ (Å) | 3.740 | 3.744 | 3.744 | 3.743 |
| $d_1$ (Å) | 4.768 | 4.768 | 4.762 | 4.776 |
| $d_2$ (Å) |  | 8.045 | 8.021 | 7.960 (= $c/2$) |

**B. Quasiparticle band structures**

Figure 2 compares the DFT and QP band structures of bulk and few-layer GaSe. Except for the band gap enlarging, the quasiparticle corrections do not seem to change the band dispersion significantly for both bulk and few-layer systems. While few-layer systems have an indirect gap, bulk GaSe has a direct gap at Γ; these result are consistent with previous calculations [33-37]. However, it should be mentioned that the nature of the band gap of bulk GaSe is still not fully settled. Experiments [38-40] seem to consistently indicate that bulk GaSe has an indirect band gap with the CBM located at the M point. However, there is also significant variation (25 – 500 meV) [38-40] in the energy difference ($\Delta_{\Gamma M} = E_\Gamma - E_M$) between the Γ and the M points. In particular, Colettia et al. [39] observed that $\Delta_{\Gamma M}$ decreases significantly with decreasing electron current. Since carriers are inevitably introduced in experiment, it is plausible the discrepancy between theory and experiment may be (at least) partially) resolved by considering the carriers renormalization effects. This issue deserves further investigation.

It is instructive to compare the band structures of the bulk and monolayer systems [Fig. 2 (a) and (b)]. The interlayer interaction, although considered weak (with an inter-layer binding energy of about 19 meV/atom), results in surprisingly large splitting of the band edge states and significant changes in band dispersions of the bulk phase compared with that of the monolayer. The valence band splitting Δ (defined in Fig. 2) increases from 0.25 eV (monolayer) to 1.13 eV (bulk), as shown in Table III.

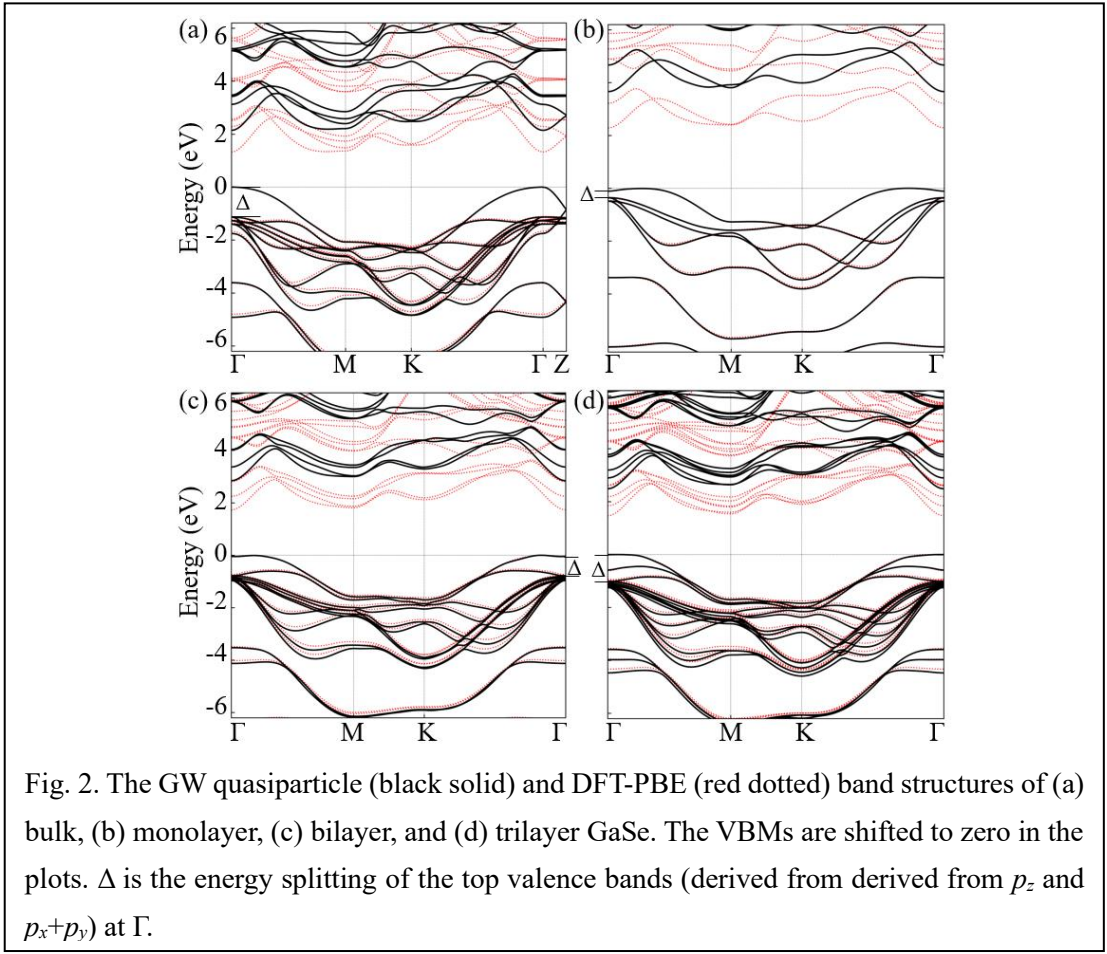

Fig. 2. The GW quasiparticle (black solid) and DFT-PBE (red dotted) band structures of (a) bulk, (b) monolayer, (c) bilayer, and (d) trilayer GaSe. The VBMs are shifted to zero in the plots. Δ is the energy splitting of the top valence bands (derived from derived from $p_z$ and $p_x+p_y$) at Γ.

One interesting observation is the dispersion of the top valence band of monolayer GaSe: the VBM is shifted away (by $\Delta k$) from $\Gamma$, resulting in a slightly indirect band gap and a MH-like band dispersion near the VBM (also shown schematically in Fig. 3). The bulk structure, on the other hand, has a VBM at $\Gamma$. The calculated parameters $\Delta k$ and $\Delta E$ (defined in the top panel of Fig. 3) for few-layer systems are listed in Table II. Both $\Delta E$ and $\Delta k$ decrease quickly as the number of layers increases, and for the bulk system, a direct gap at the $\Gamma$ point is formed. The MH-like dispersion also gives an extraordinary sharp density of states (DOS) near the VBM for the monolayer system, as shown in the bottom panel of Fig. 3. The decomposition of the DOS indicates that the top valence band has predominantly the Se-$p_z$ character.

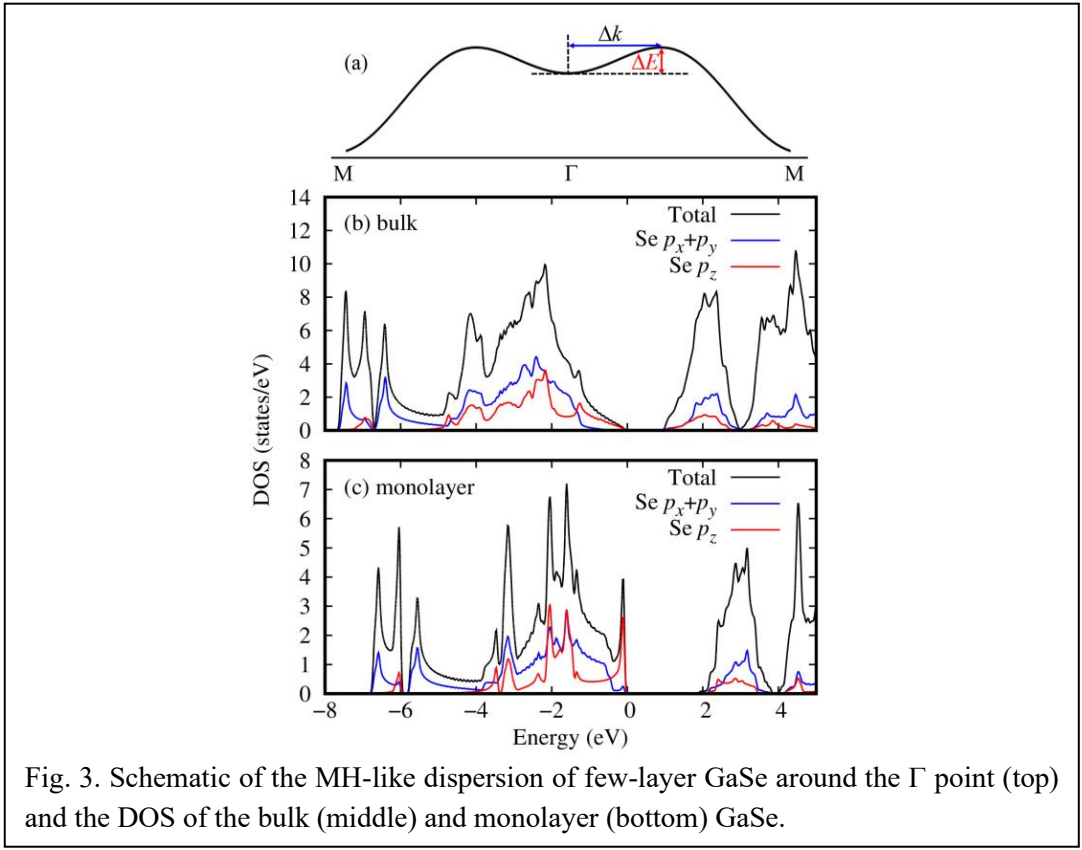

Fig. 3. Schematic of the MH-like dispersion of few-layer GaSe around the $\Gamma$ point (top) and the DOS of the bulk (middle) and monolayer (bottom) GaSe.

The interlayer $p_z$-$p_z$ coupling greatly suppresses this MH-like dispersion and the DOS near VBM as the number of layers increases. These interesting features have been discussed in great detail previously [10]. The Se-$p_x$+$p_y$ component (colored blue in the middle and bottom panels of Fig. 3), on the other hand, is minimally affected by the interlayer coupling. The large $\Delta E$ and $\Delta k$ for the monolayer and their layer-dependence explain the significant decrease in PL emission intensity with a decreasing number of layers and the disappearance of PL signals in monolayer GaSe [15]. PL from the bilayer and trilayer systems (although much weaker compared with that from bulk GaSe) is likely due to thermal population of the excited carriers since the direct-indirect energy difference $\Delta E$ is comparable or smaller than the thermal energy at room temperature. Other mechanisms that may contribute to the PL in the bilayer and trilayer GaSe include phonon-assisted PL [41,42] or and exciton dispersion effects [43].

Table II. Parameters $\Delta E$ (meV) and $\Delta k$ (Å$^{-1}$) (defined in the top panel of Fig. 3) for the MH-like dispersion of the top valence band of few-layer GaSe.

|  | 1L | 2L | 3L |
|---|---|---|---|
| $\Delta E$ | 128 | 47 | 18 |
| $\Delta k$ | 0.17 | 0.11 | 0.06 |

Table III compares the calculated (DFT and GW) band gap with those measured by various methods. For few-layer systems, we list both the direct gap at the $\Gamma$ point and the minimum gap, calculated within DFT and GW methods. The calculated minimum quasiparticle band gaps ($E_{g,\text{min}}^{\text{GW}}$ in Table III, highlighted in bold) of few-layer systems agree reasonably well (< 0.2 eV) with those measured by scanning tunneling spectroscopy [$E_g^{\text{QP}}(\text{exp})$ in Table III, highlighted in bold] [17]. Band gaps measured by optical techniques (e.g., CL or PL) [$E_g^{\text{opt}}(\text{exp})$ in Table III, highlighted in bold] should be compared with the excitonic gaps ($E_g^{\text{ex}}$ in Table III, to be discussed later), more specifically, the calculated bright excitonic gap highlighted in bold. For example, the calculated minimum excitonic gap of monolayer GaSe is 3.09 eV. However, this value cannot be compared directly with experimentally measured optical gap since the first exciton is a dark exciton. The calculated bright excitonic gap for monolayer GaSe is 3.39 eV, which compares well with the experimental result of 3.3 eV. For the bilayer system, the first exciton state is bright (albeit weak). For bulk GaSe, since the exciton binding energy is small, the measured optical gap agrees well with the calculated GW gap.

We mention that GW results calculated using different GPP models (or without the use of GPP models) can sometimes differ significantly. In particular, it has been shown that the HL-GPP model often gives larger band gap compared with other models [44,45]. However, the accuracy of GW calculations is a fairly complex issue. In addition to various approximations made in the calculations, numerical convergence is also an important issue [26,27,46-48]. Detailed discussion of these issues is beyond the scope of this work. In the next two sections, we discuss the excitonic structure and optical properties in more detail.

Table III. Quasiparticle and optical band gaps (in eV) of few-layer and bulk GaSe: Comparison between theory and experiment. Δ is the GW-predicted valence band splitting (in eV) at Γ, as defined in Fig. 3.

|  | 1L | 2L | 3L | Bulk |
|---|---|---|---|---|
| $E_{g,\Gamma}^{PBE}$ | 2.26 | 1.65 | 1.36 | 1.17 |
| $E_{g,min}^{PBE}$ | 2.13 | 1.60 | 1.34 | 1.17 |
| $E_{g,\Gamma}^{GW}$ | 3.77 | 2.89 | 2.49 | 2.14 |
| $E_{g,min}^{GW}$ | **3.64** | **2.84** | **2.47** | **2.14** |
| $E_g^{QP}$ (exp) | 3.5[1] | 3.00[1] | 2.30[1] | 2.13[2] |
| $E_g^{ex}$ | 3.09 (dark) **3.39 (bright)** | **2.49 (weak)** | 2.20 (dark) **2.32 (weak)** | **2.10 (bright)** |
| $E_g^{opt}$ (exp) | 3.3[3] | 2.44[4] | 2.28[4] | 1.99[4], 2.11[5] |
| Δ | 0.25 | 0.74 | 1.02 | 1.13 |

[1]STS [17]; [2]optical absorption [49,50]; [3]CL [16]; [4]PL peak positions [15]; [5]optical absorption [49-51]

## C. Excitonic structure and optical properties of bulk GaSe

Several experiments have been conducted to study the optical properties of bulk [49-53] and few-layer [15,16,49-54] GaSe. However, optical properties calculations have only been reported for monolayer GaSe [37,55,56], and there is a lack of systematic studies of layer-dependent excitonic structure and optical properties of these systems. In this work, the excitonic structure and optical properties of few-layer and bulk GaSe are investigated by solving the BSE [19], which is reduced to an eigenvalue problem within the Tamm-Dancoff approximation:

$$(E_{c\mathbf{k}} - E_{v\mathbf{k}})A_{vc\mathbf{k}}^S + \sum_{v'c'\mathbf{k}'} \langle vc\mathbf{k}|K^{eh}|v'c'\mathbf{k}'\rangle A_{v'c'\mathbf{k}'}^S = \Omega^S A_{vc\mathbf{k}}^S, \quad (1)$$

where $E_{c\mathbf{k}}$ and $E_{v\mathbf{k}}$ are the quasiparticle energies of the conduction and valence states; $A_{vc\mathbf{k}}^S$ is the eigenvector (i.e., the *e-h* amplitude) that defines the exciton wave function:

$$\Psi^S(\mathbf{r}_e, \mathbf{r}_h) = \sum_{vc\mathbf{k}} A_{vc\mathbf{k}}^S \psi_{c\mathbf{k}}(\mathbf{r}_e)\psi_{v\mathbf{k}}^*(\mathbf{r}_h), \quad (2)$$

and $K^{eh}$ is the *e-h* interaction kernel. Solving the above eigenvalue problem gives the *e-h* excitation energy $\Omega^S$, and the wave function of the excitonic state $|S\rangle$. The imaginary part of the

macroscopic transverse dielectric function is

$$\epsilon_2(\omega) = \frac{16\pi^2 e^2}{\omega^2} \sum_S |\vec{e} \cdot \langle 0|\vec{v}|S\rangle|^2 \delta(\omega - \Omega^S), \qquad (3)$$

where $\vec{v}$ is the velocity operator, $\vec{e}$ is the polarization vector of the light, $\omega$ is the energy of the photon.

Bulk GaSe is known to have a highly anisotropic optical response [49-53]: strong excitonic absorption is observed near the band gap (2.11 eV at low temperatures) for photon polarization (*E*) direction parallel to the *c*-axis (*E*//*c*); for *E* parallel to the *ab*-plane (*E*//*ab*), however, no significant optical activities are observed below 3.0 eV. Our theoretical findings, as presented in Fig. 4, are consistent with the experiment. In particular, the calculated first excitonic absorption peak for *E*//*c* at 2.10 eV (upper panel of Fig. 4) agrees very well with the low-temperature experimental results of about 2.11 eV [49-51]. In the upper panel of Fig. 4, we also show the energies of excitonic states as vertical lines scaled by the dipole transition matrix elements for polarization *E*//*c*. The strong absorption peak for *E*//*ab* at about 3.70 eV (lower panel of Fig. 4) also agrees well with the experimental value of 3.63 eV [52].

In addition to the optical absorption shown in Fig. 4, understanding the strength of *e-h* interaction and the exciton binding energies ($E_b^{ex}$) of the excitons is also of great interest. The exciton binding energy is often loosely defined as the difference between the quasiparticle excitation gap and the excitonic gap, i.e., $E_b^{ex,S} = E_g^S - \Omega^S$. The quasiparticle gap $E_g^S$, however, must be understood as the weighted average of the non-interacting *e-h* excitation gap [57,58], defined as, for any excitonic state $|S\rangle$,

$$E_g^S = \sum_{vc\mathbf{k}} |A_{vc\mathbf{k}}^S|^2 (E_{c\mathbf{k}} - E_{v\mathbf{k}}). \qquad (4)$$

In the bottom panel of Fig. 4, we show the calculated binding energies (blue dots in the figure) for all excitonic states below 6 eV for bulk GaSe. Most states have a binding energy smaller than 50 meV. However, a few excitonic states that are well above the fundamental band gap (at about 3.49 eV) have surprisingly large binding energies, with one state having a binding energy as large as 168 meV. The dramatic variation of the exciton binding energy from state to state and the abnormally large binding energies for some exciton states in such a moderate-gap bulk semiconductor deserve scrutiny. Although the binding energy of an excitonic state is generally related to the spread (or localization) of the exciton wave function, it is not yet well understood why some states are more localized than others in a single system.

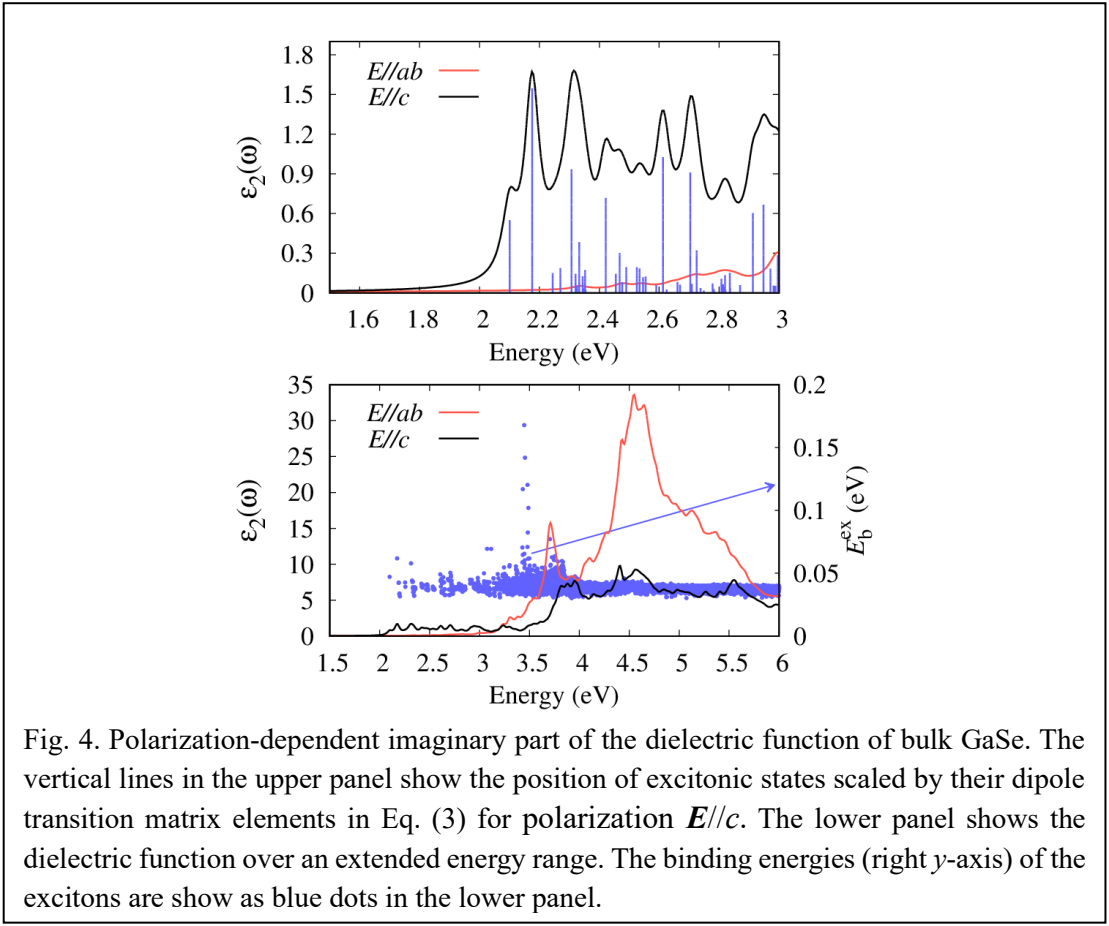

Fig. 4. Polarization-dependent imaginary part of the dielectric function of bulk GaSe. The vertical lines in the upper panel show the position of excitonic states scaled by their dipole transition matrix elements in Eq. (3) for polarization $\boldsymbol{E}//c$. The lower panel shows the dielectric function over an extended energy range. The binding energies (right $y$-axis) of the excitons are show as blue dots in the lower panel.

To this end, we investigate the distribution (spread) of the exciton wave functions in both the momentum and real spaces. First, we define the band- and $k$-resolved hole and electron amplitudes,

$$\left|A_{v\boldsymbol{k}}^{S}\right|^{2} = \sum_{c}\left|A_{vc\boldsymbol{k}}^{S}\right|^{2}, \tag{5}$$

$$\left|A_{c\boldsymbol{k}}^{S}\right|^{2} = \sum_{v}\left|A_{vc\boldsymbol{k}}^{S}\right|^{2}, \tag{6}$$

and the $k$-resolved pair amplitude,

$$\left|A_{\boldsymbol{k}}^{S}\right|^{2} = \sum_{vc}\left|A_{vc\boldsymbol{k}}^{S}\right|^{2}. \tag{7}$$

These functions reveal the distribution of the hole, electron, and the $e$-$h$ pair in the $k$-space of a given state $|S\rangle$. The top panels of Fig. 5 show $\left|A_{v\boldsymbol{k}}^{S}\right|^{2}$ and $\left|A_{c\boldsymbol{k}}^{S}\right|^{2}$ of two representative excitonic states, A and B. Whereas exciton A (top left panel) has a calculated binding of about 47 meV, that of exciton B (top right panel) is 168 meV. The electron and hole amplitudes of exciton A are highly

concentrated in a small region of the BZ (in this case, near the Γ point). In contrast, for exciton B, these amplitudes are rather delocalized in the momentum space over which there is a pair of valence and conduction bands that are nearly parallel in portion of the BZ. It is these (nearly) parallel valence and conduction bands that offer a large number of *e-h* pairs across the BZ with similar excitation energies, giving rise to excitonic states that are delocalized in the momentum space.

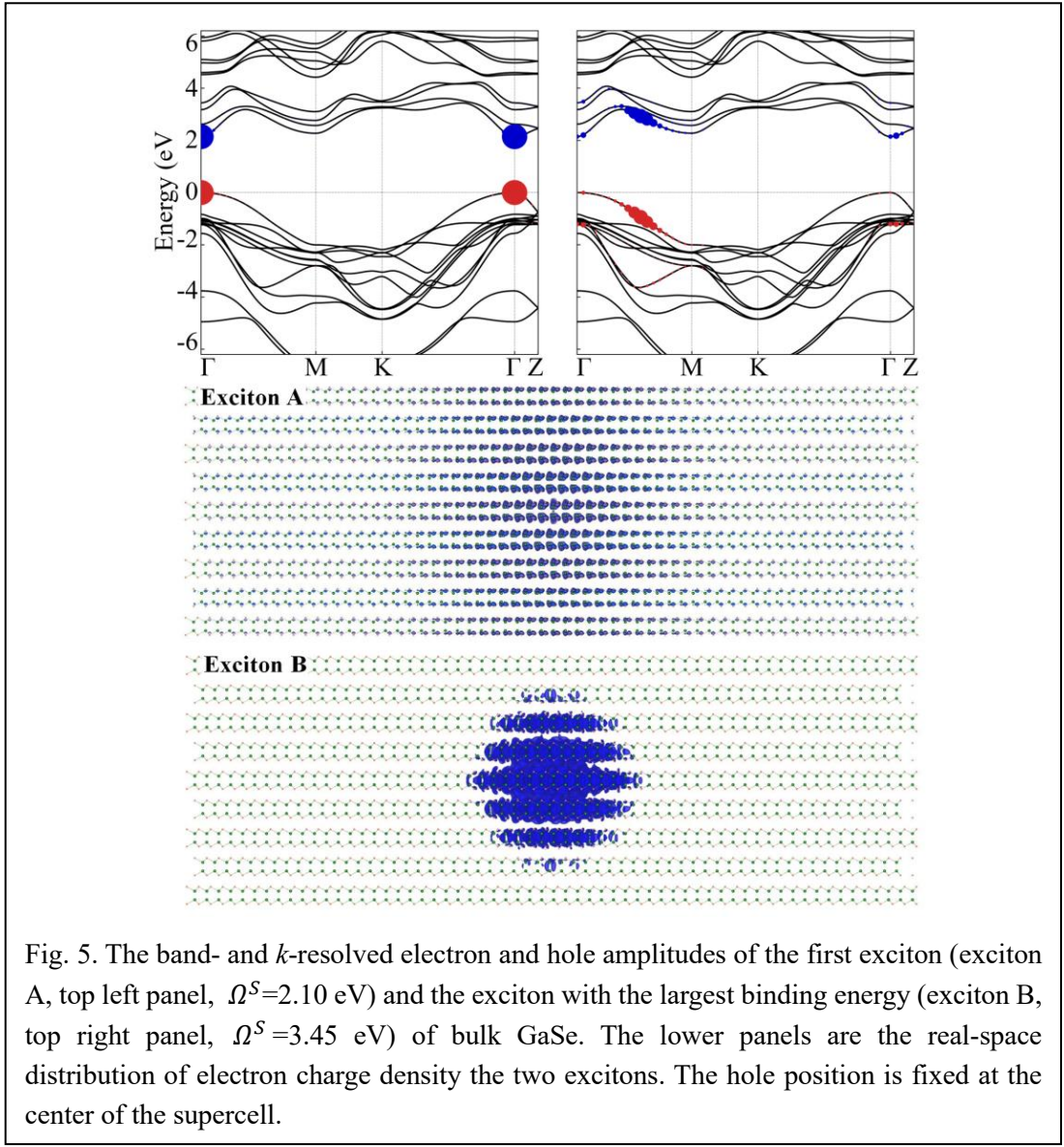

Fig. 5. The band- and *k*-resolved electron and hole amplitudes of the first exciton (exciton A, top left panel, $\Omega^S$=2.10 eV) and the exciton with the largest binding energy (exciton B, top right panel, $\Omega^S$ =3.45 eV) of bulk GaSe. The lower panels are the real-space distribution of electron charge density the two excitons. The hole position is fixed at the center of the supercell.

The localization (or delocalization) of the electron and hole amplitudes in the momentum space is highly correlated with the spread of the exciton wave function in the real space. In the bottom two panels of Fig. 5, we plot the isosurface of the square of the electron part of the exciton wave function [defined as $\rho_e(r_e; r_h) = |\Psi^S(r_e, r_h)|^2 / \int |\Psi^S(r_e, r_h)|^2 dr_e$] by fixing the hole position $r_h$ near the center of the cell. For exciton A, the electron wave function is rather delocalized. This

is in sharp contrast with that of exciton B, for which the electron wave function is highly localized near the hole. The existence of such strongly localized (strongly bound, thus potentially long lifetime) excitons in the excitation continuum might have interesting implications in the dynamics of excitons and deserves further investigations. We would like to mention that the distribution of the electron density does not depend sensitively on the choice of the hole position; see Supplemental Material at [30] Fig. S3 for the plots of the electron density distribution with difference choices of the hole position.

To gain a more quantitative understanding, we define a measure for the spread of the electron, $r_e = \langle r_e^2 \rangle^{-1/2}$; for a given excitonic state $|S\rangle$ and a fixed hole position $r_h$,

$$\langle r_e^2 \rangle = \int |\Psi^S(r_e, r_h)|^2 (r_e - r_h)^2 dr_e \Big/ \int |\Psi^S(r_e, r_h)|^2 dr_e. \tag{7}$$

We expect that the exciton binding energy would inversely proportional to the size of the exciton in real space. Indeed, our calculations of about 20 excitons suggest a nearly linear relationship between the exciton binding energy and $1/r_e$ as shown in Fig. 6. It should be mentioned that the above definition of the electron radius may appear rather arbitrary since it depends on the choice of the hole position $r_h$. However, this is not the case. In fact, the electron radius defined this way is independent of the choice of the hole position (within numerical errors); see Table S1 in Supplemental Material at [30] for the effective electron radius of the exciton of bulk GaSe with largest binding energy calculate with different choices of the hole position.

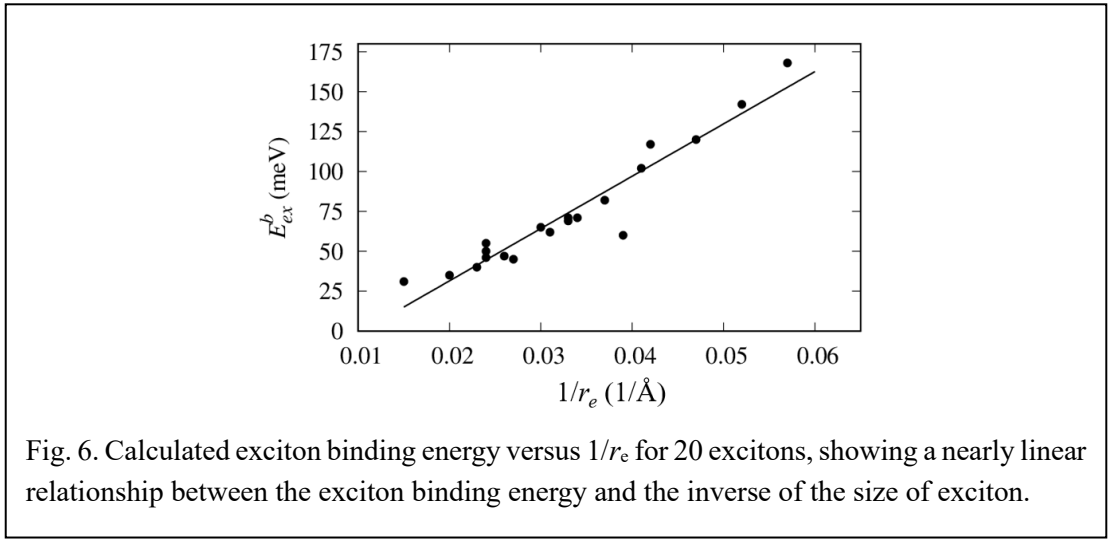

Fig. 6. Calculated exciton binding energy versus $1/r_e$ for 20 excitons, showing a nearly linear relationship between the exciton binding energy and the inverse of the size of exciton.

It should be mentioned that it is extremely difficult to carry out fully converged BSE calculations for the excitonic properties, especially the binding energies, of delocalized excitons. The measured binding energy of the lowest excitonic state in bulk GaSe ranges from 20 to 22 meV [50,53,59], to be compared with the theoretical value of 47 meV calculated using a 36×36×8 $k$- grid. Although such a k-grid density is sufficient to converge the overall features of the optical absorption spectra shown in Fig. 4. Additional details can be found in Fig. S2 of Supplemental Material at [30]. The calculated binding energies of most excitonic states (except for the localized states discussed above) are hardly converged with respect to the k-grid density. Unfortunately, exciton calculations using k-grid densities higher than 36×36×8 for this system are beyond our local computing capacity. In fact, even with a 36×36×8 k-grid, the e-h Hamiltonian matrix size is already about 500,000 × 500,000 in this case.

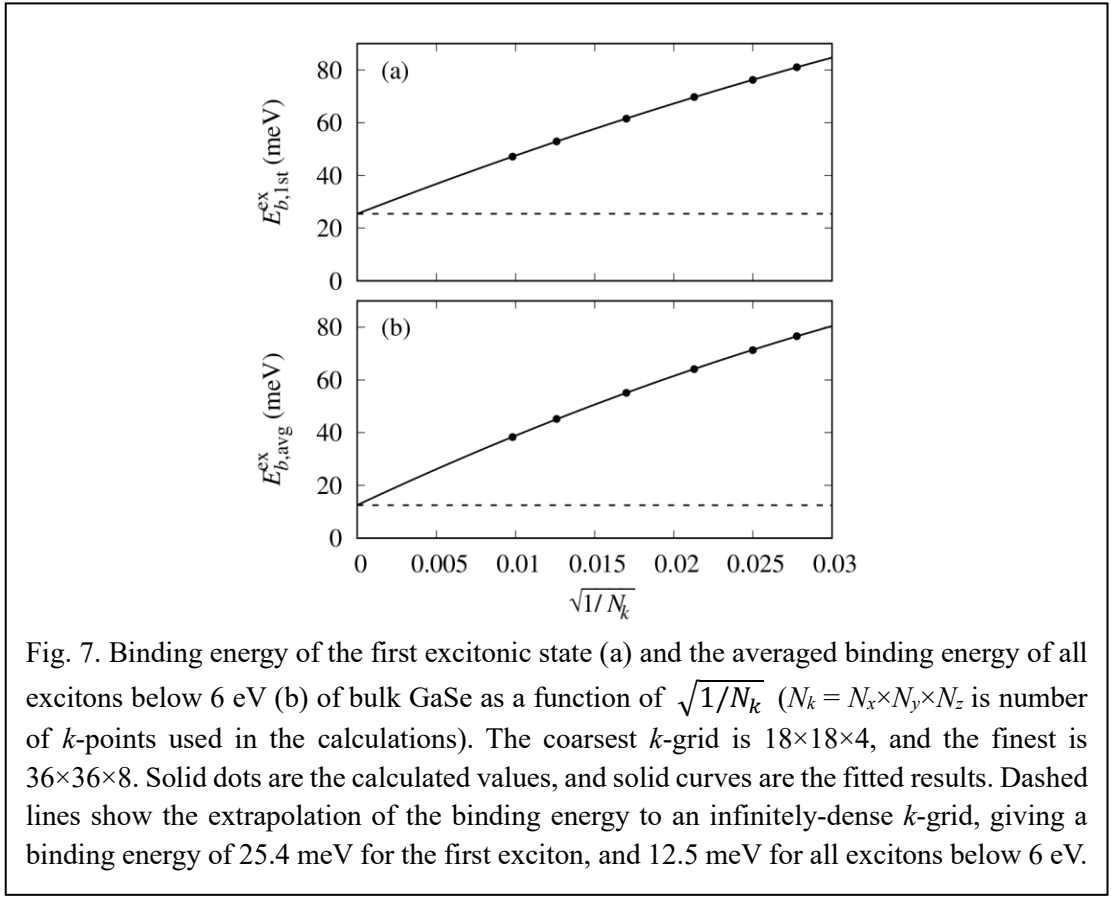

Fig. 7. Binding energy of the first excitonic state (a) and the averaged binding energy of all excitons below 6 eV (b) of bulk GaSe as a function of $\sqrt{1/N_k}$ ($N_k = N_x \times N_y \times N_z$ is number of k-points used in the calculations). The coarsest k-grid is 18×18×4, and the finest is 36×36×8. Solid dots are the calculated values, and solid curves are the fitted results. Dashed lines show the extrapolation of the binding energy to an infinitely-dense k-grid, giving a binding energy of 25.4 meV for the first exciton, and 12.5 meV for all excitons below 6 eV.

Figure 7 shows the convergence behavior of the binding energy of the first exciton and the averaged binding energy of all excitonic states below 6.0 eV. It is interesting to note that the binding energies scale approximately linearly with $\sqrt{1/N_k}$, where $N_k = N_x \times N_y \times N_z$ is the total number of k-points used in expanding the exciton wave functions [see Eq. (1)]. Since the volume over which the exciton wave functions are represented is $V=N_k \times V_{cell}$, an enormously large cell (i.e., a very high k-grid density) is needed to converge the properties of delocalized excitons in bulk semiconductors. The extrapolated exciton binding energy to an infinitely dense k-grid is 25.4 meV for the first exciton, which agrees well with the experimental value of 20~22 meV [50,53,59].

## D. Layer-dependent excitonic structure and optical properties

Next, we investigate the effects of layer thickness on the excitonic structure and optical properties of GaSe. It is instructive to compare $\varepsilon_{00}^{-1}(q)$ (the $q$-dependent head element of the inverse dielectric matrix) of monolayer, multilayer, and bulk systems, as shown in Fig. 8, where $q$ is along the $ab$ plane. The rapid upshift of this function at small $q$ (approaching 1 as $q$ approaches 0) is characteristic of 2D materials. This behavior also explains the extremely slow convergence of the calculated quasiparticle properties with respect to the BZ sampling $k$-grid density unless special sampling techniques [27,60,61] are used. With the increasing layer thickness, the dielectric screening $[1/\varepsilon_{00}^{-1}(q)]$ gradually increases. However, even for the trilayer system (the thickness of the trilayer GaSe is about 21 Å), the dielectric screening is still significantly reduced compared with that of the bulk material.

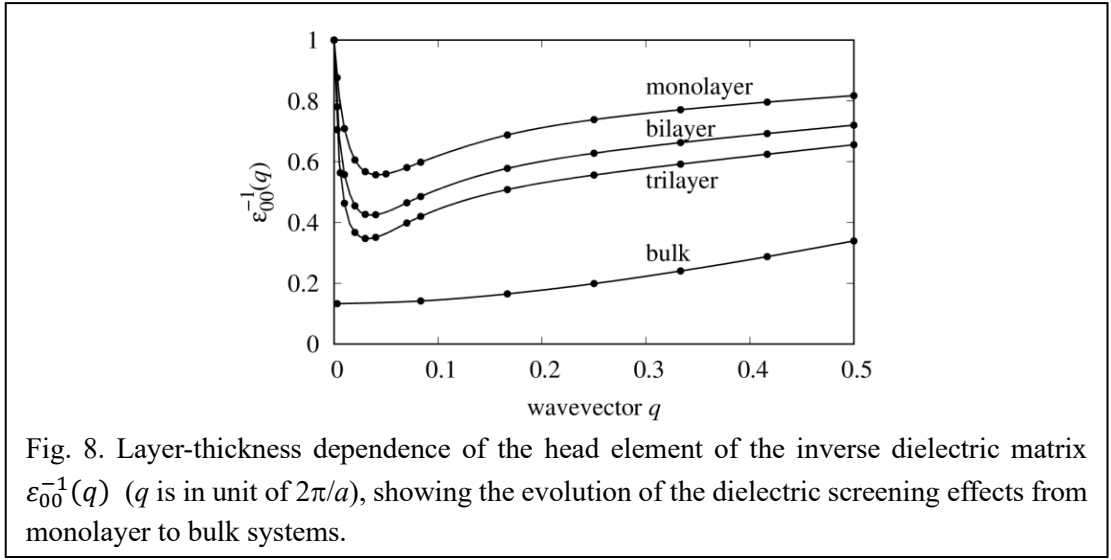

Fig. 8. Layer-thickness dependence of the head element of the inverse dielectric matrix $\varepsilon_{00}^{-1}(q)$ ($q$ is in unit of $2\pi/a$), showing the evolution of the dielectric screening effects from monolayer to bulk systems.

The layer-thickness-dependent dielectric screening and the interlayer coupling will significantly affect the excitonic structure and optical absorption of these systems. Figure 9 compares the calculated optical absorption spectra (for **E**//$ab$, black curves) and excitonic structure (exciton binding energy vs. exciton energy, shown in blue dots) of few-layer and bulk GaSe. The vertical black dashed line indicates the direct quasiparticle gap $E_g$ at $\Gamma$; the shaded areas around $E_g$ highlight the low energy excitonic states of each system, which are then expanded in the narrow horizontal boxes atop each main plot to show their optical activities. Each vertical line in the narrow horizontal boxes indicates an excitonic state, color-coded according to its optical dipole matrix element. Note that for 2D materials, the absolute value of calculated $\varepsilon_2$ bears no significance since it depends on the volume of the calculation unit cell (more precisely, the $c$ lattice constant). However, the main features of the absorption spectra can be directly compared with the experiment.

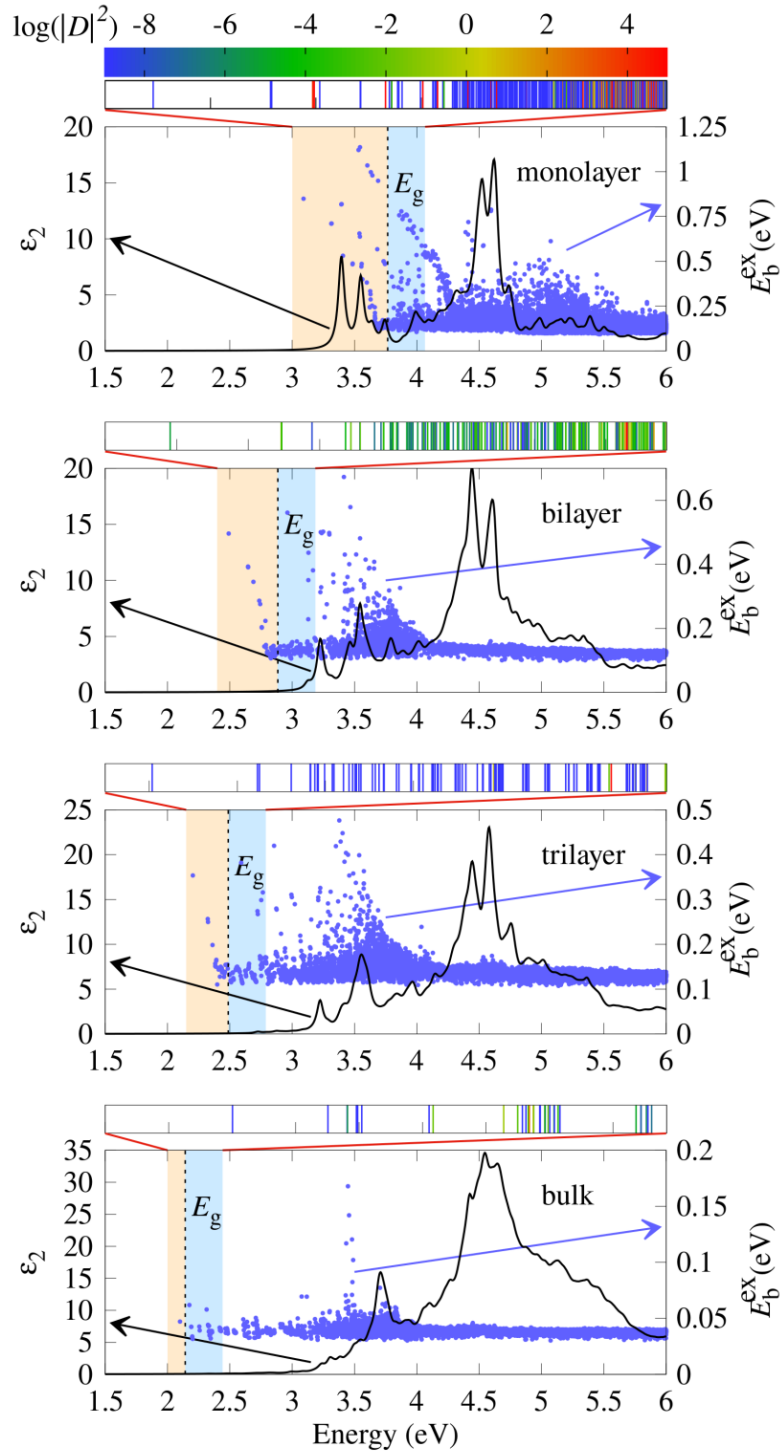

Fig. 9. Optical absorption spectra (**E**//*ab*, black curves) and exciton binding energies $E_b^{ex}$ (blue dots) of few-layer and bulk GaSe. The vertical black dash line is the position of QP band gap ($E_g$) at Γ. An energy broadening of 0.03 eV is used in the calculation of ε$_2$. The shaded areas highlight the excitonic states slightly below (light orange area) and above (light blue) the fundamental gap. The low energy excitonic states within the shaded area, color-coded with their dipole matrix elements, are shown in the extended horizontal boxes on top of the main plots.

Low-energy excitonic states and optical properties are highly layer-thickness dependent, and the systematic changes in the first two excitonic absorption peaks can be clearly seen as the layer thickness increases. Although all few-layer systems have a large number of below-the-gap (BTG) excitonic states, the number of these BTG excitons decreases significantly with increasing layer thickness as a result of reduced *e-h* interaction. For bulk GaSe, we only observe one excitonic state below the fundamental gap. It is possible that one needs higher density *k*-grids to resolve more excitonic Rydberg states below the band gap. Another interesting observation is that only the monolayer shows strong optical absorption below the band gap.

The first absorption peak is very strong for monolayer GaSe, whereas in the few-layer and bulk systems, it is the second peak that is more pronounced. For the monolayer system, the first absorption peak can be attributed to a pair of excitons formed by the Se $p_x+p_y$ derived hole states (colored red in the left panel of Fig. 10) and primarily Ga *s*-derived electron states (colored blue in the left panel of Fig. 10) near the $\Gamma$ point, whereas the second absorption peak is associated with a single exciton mainly composed of $p_z$ derived, nearly parallel valence and conduction bands midway along the $\Gamma$ to M path. These features are similar to those shown in Fig. 5 in the case of bulk GaSe. See Fig. S4 in Supplemental Materials at [30] for plots of the *e-h* pair amplitudes [Eq. (7)] of a few selected excitonic states.

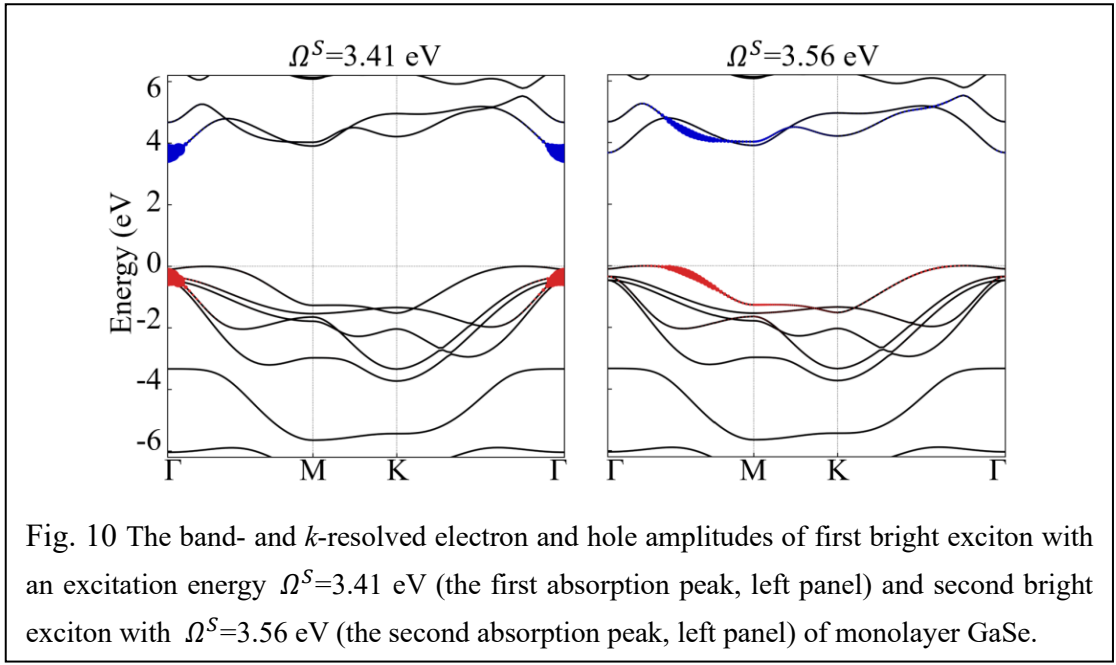

Fig. 10 The band- and *k*-resolved electron and hole amplitudes of first bright exciton with an excitation energy $\Omega^S=3.41$ eV (the first absorption peak, left panel) and second bright exciton with $\Omega^S=3.56$ eV (the second absorption peak, left panel) of monolayer GaSe.

Another interesting observation of the plot of exciton binding energy vs excitation energy is that, on top of the seemingly random distribution of the exciton binding energy, the binding energy of some states seem to group into a narrow, shell-like distribution. These structures arise from the presence of nearly parallel valence and conductions bands as we have discussed in great detail in our recent works [57,58].

While the changes in the overall absorption spectra from the monolayer to bilayer are highly significant (suggesting the strong effects from the enhanced screening and interlayer coupling), those from bilayer GaSe to bulk are relatively minor. This is somewhat surprising considering that the excitonic structure (positions and distribution of the excitons, their binding energies, etc.) are very different in these systems (i.e., bilayer, trilayer, and bulk) as can be seen from Fig. 9. The strong absorption peak around 4.5 eV is fairly robust across all systems, indicating that optical absorption near this energy is dominated by intra-layer excitons.

The calculated exciton binding energies decrease systematically with increasing layer thickness. For bulk GaSe, most excitonic states have binding energies smaller than 50 meV, with a few states having noticeably large binding energy states around 3.5 eV, as discussed in Section III C. Table IV shows the minimum quasiparticle band gaps, excitation energies and binding energies of the first exciton and the exciton with the largest binding energy of few-layer and bulk GaSe. We also show the binding energy of the first exciton using the conventional definition, i.e., i.e., $E_{g,min}^{GW} - \Omega^1$ in parentheses for comparison. We would like to point out that this definition is only valid for the excitons for which the electron and hole states are derived from QP states that define the fundamental band gap, which is often not the case.

Table IV. Quasiparticle band gaps and exciton energies of few-layer and bulk GaSe. Only the first exciton and the exciton with the largest binding energy are shown. The exciton binding energies are calculated using the quasiparticle gap defined in Eq. (4). Exciton binding energies calculated using the conventional definition (i.e., $E_{g,min}^{GW} - \Omega^1$) are shown in parentheses. The $E_b^1$ for bulk GaSe is obtained by extrapolating theoretical results to an infinitely-dense $k$-grid.

|  | 1L | 2L | 3L | Bulk |
|---|---|---|---|---|
| $E_{g,min}^{GW}$ | 3.64 | 2.84 | 2.47 | 2.14 |
| $\Omega^1$ (eV) | 3.09 | 2.49 | 2.20 | 2.10 |
| $E_b^1$ (meV) | 849 (550) | 493 (350) | 354 (270) | 25 (25) |
| $\Omega^{max}$ (eV) | 3.54 | 3.42 | 3.38 | 3.45 |
| $E_b^{max}$ (meV) | 1136 | 673 | 476 | 168 |

The excitonic gap decreases monotonically with the increasing layer thickness. The binding energy decreases by over 350 meV (or more than 40%) going from the monolayer to the bilayer system. Interestingly, the excitonic gap of the trilayer is already very close to that of the bulk. However, the binding energy of the first exciton of trilayer GaSe is still more than 10 times larger than that of bulk GaSe. The largest binding energy increases from 168 meV (bulk) to 1.14 eV

(monolayer), as shown in Table IV. We mention that the excitonic gap (i.e., the energy of the first exciton) may or may not correspond to the measured optical gap depending on the optical dipole matrix element of the transition. For example, the first exciton is dark for the few-layer systems. The calculated layer-dependent optical gaps, as measured by the first bright exciton energies, agree very well with the experiment (see Table III).

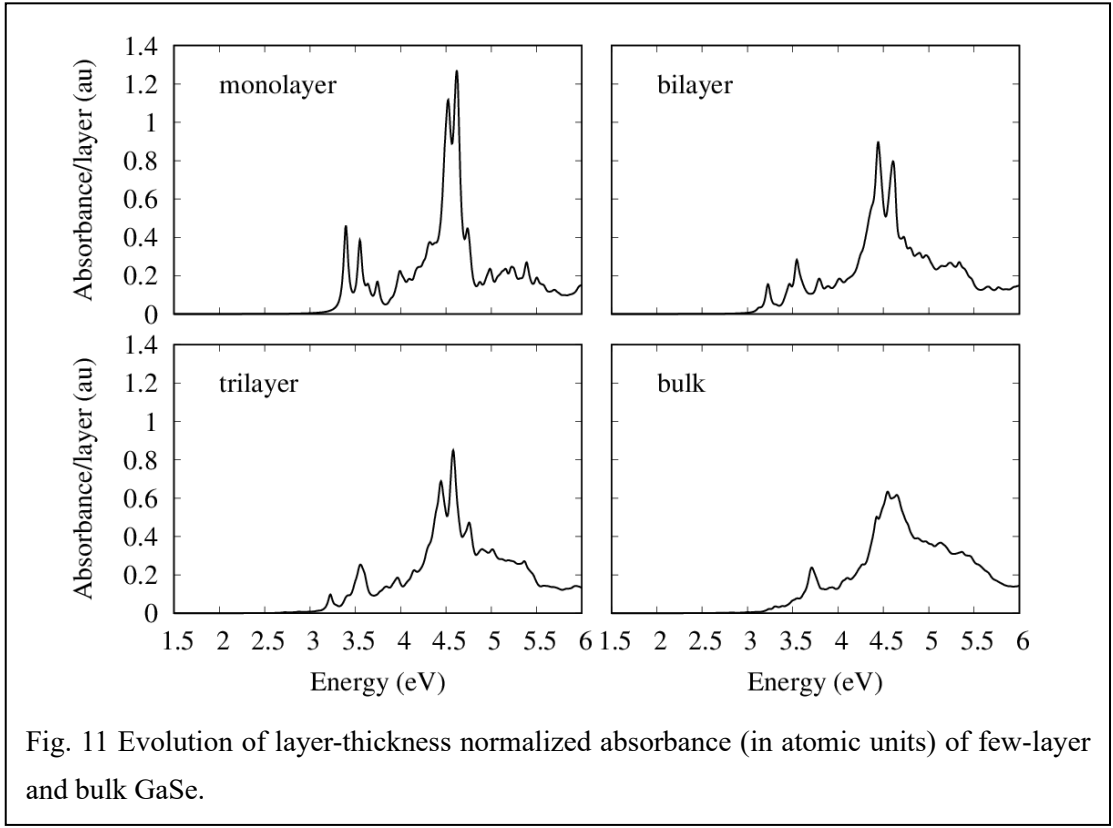

Fig. 11 Evolution of layer-thickness normalized absorbance (in atomic units) of few-layer and bulk GaSe.

Finally, it should be mentioned that for 2D materials, $\varepsilon_2$ depends on the thickness of the vacuum layer included in the calculations. Therefore, the absolute values of $\varepsilon_2$ for 2D materials bear no significance and cannot be compared directly with that of bulk. However, the evolution of the excitonic properties (i.e., the position and relative strength of the excitonic absorption peaks, and the exciton binding energies) shown in Fig. 9 can be compared directly with experiment. To gain better understanding of the evolution of the optical absorption, we shown in Fig. 11 layer-normalized absorbance [62], defined as $A(\omega) = L \cdot \omega/c \cdot \epsilon_2(\omega)/N$, where $L$ is the lattice constant along the $c$-axis, and $N$ is the number of layers (1 for monolayer GaSe, 2 for bilayer, 3 for trilayer, and 2 for bulk). This quantity better shows the evolution of the absorption efficiency with layer thickness. Overall, the monolayer shows highest absorption efficiency (per layer).

## IV. Summary

We have performed detailed GW plus BSE calculations for few-layer and bulk GaSe. The unique excitonic structures and exciton binding energies of these systems are carefully analyzed to illustrate the effects of interlayer coupling and layer thickness dependent dielectric screening on excited state properties of GaSe. Our results also help resolve apparent discrepancies between different experiments. The interlayer coupling greatly suppresses the Mexican-hat-like dispersion of the top valence band, explaining the greatly enhanced PL as layer thickness increases. We find the delocalization (localization) of excitonic wave functions in the BZ results in their localization (delocalization) in the real-space and increased (decreased) exciton binding energies. The exciton binding energy is inversely proportional to spread of the averaged *e-h* separation (i.e., spread of the exciton wave function). The presence of (nearly) parallel valence and conduction bands greatly facilitates the delocalization of exciton wave functions in the BZ, leading strongly localized wave function in the real space, and thus greatly enhanced exciton binding energy. The calculated maximum exciton binding energy decreases from 1.14 eV (monolayer) to 0.17 eV (bulk). For delocalized excitons, an extremely dense *k*-grid may be needed to converge the calculated exciton binding energy. The calculated exciton binding energy of the lowest excitonic state of bulk GaSe, after extrapolating to the converged value, agrees very well with experimental result. The changes in the overall absorption spectra from the monolayer to the bilayer are highly significant, and the monolayer is the only case that exhibits bright exciton absorption (for polarization **E**//*ab*) below the fundamental gap.

**Acknowledgments**

Work at HDU and SHU was supported by National Natural Science Foundation of China (11929401, 12074241, 52120204) and Key Research Project of Zhejiang Lab (2021PE0AC02). W.G. was supported by National Natural Science Foundation of China (12104080). Work at UB was supported by the National Science Foundation under Grants No. DMREF-1626967. We acknowledge computational support from the Center for Computational Research, University at Buffalo, SUNY.